\documentclass[twoside,final]{pro12}

\usepackage[ansinew]{inputenc}
\usepackage{amsmath}
\usepackage{graphicx}
\usepackage{array}
\usepackage{lscape}
\usepackage{multirow}
\usepackage{siunitx}
\usepackage{float}
\usepackage{natbib}
\usepackage{hyperref}

\head{D.\,M. Weigt and L.\,A. Ca$\tilde{n}$izares et al.}{The SURROUND Mission}	

\begin{document}

\title{The science behind SURROUND: a constellation of CubeSats around the Sun}	
\author{D. M. Weigt$^\dagger$\adress{\textsl School of Cosmic Physics, Dublin Institute for Advanced Studies, Dublin, Ireland}\,\,$^,$\adress{\textsl Department of Computer Science, Aalto University, Aalto, Finland}$\,\,$, L. A. Ca$\tilde{\mathrm{n}}$izares$^{\dagger1}$$^,$\adress{\textsl School of Physics, Trinity College Dublin, Dublin , Ireland}$\,\,$, S. A. Maloney$^1$, S. A. Murray$^1$,\\ E. P. Carley$^1$, P. T. Gallagher$^1$, A. Macario-Rojas\adress{\textsl School of Engineering, The University of Manchester, Manchester, UK}$\,\,$, N. Crisp$^4$ and C. McGrath$^4$\\
\footnotesize $^\dagger$these authors contributed equally to this work}
% Here Author4 has the same affiliation as Author3

\maketitle

\begin{abstract}
One of the greatest challenge facing current space weather monitoring operations is forecasting the arrival of coronal mass ejections (CMEs) and Solar Energetic Particles (SEPs) within their Earth-Sun propagation timescales. Current campaigns mainly rely on extreme ultra-violet and white light observations to create forecasts, missing out many potential events that may be hazardous to Earth's infrastructure undetectable at these wavelengths. Here we introduce the SURROUND mission, a constellation of CubeSats each with identical radio spectrometers, and the results of the initial Phase-0 study for the concept. The main goal of SURROUND is to monitor and track solar radio bursts (SRBs), widely utilised as a useful diagnostic for space weather activity, and revolutionise current forecasting capabilities. The Phase-0 study concludes that SURROUND can achieve its mission objectives using 3 - 5 spacecraft using current technologies with feasible SEP and CME forecasting potential: a first for heliospheric monitors.

\end{abstract}

\section{The SURROUND Concept}

Observing, analysing and forecasting potentially hazardous space weather phenomena such as coronal mass ejections (CMEs: e.g.,  \citet{webb2012cmes}) and solar flares (e.g., \citet{fletcher2011flares}), have received increased attention in recent years. For example, CMEs impacting the terrestrial space environment can perturb and even compress the Earth's geomagnetic field pushing the radiation belts closer towards the planet whereas accelerated charged particles from previous flaring activity have the potential to damage/blind orbiting spacecraft. The potential impact of these events on our technologies and infrastructure (e.g., GPS blackouts, overloaded power grids...) is the main cause of this surge in research resulting many governments across the world to recognise the effects of extreme space weather as a risk.\footnote{For example, the UK government have included `severe space weather' as a `Significant' risk in their \href{https://assets.publishing.service.gov.uk/government/uploads/system/uploads/attachment_data/file/1175834/2023_NATIONAL_RISK_REGISTER_NRR.pdf}{2023 Risk register}, an increase from `Moderate' as published in their 2020 report.} Therefore many have invested in operations and monitoring to track potentially hazardous events with the view that it is imperative that forecasting and nowcasting such events are greatly improved, to potentially avoid permanent damage to society.

\begin{figure}
	\centering
	\noindent\includegraphics[width=0.55\textwidth]{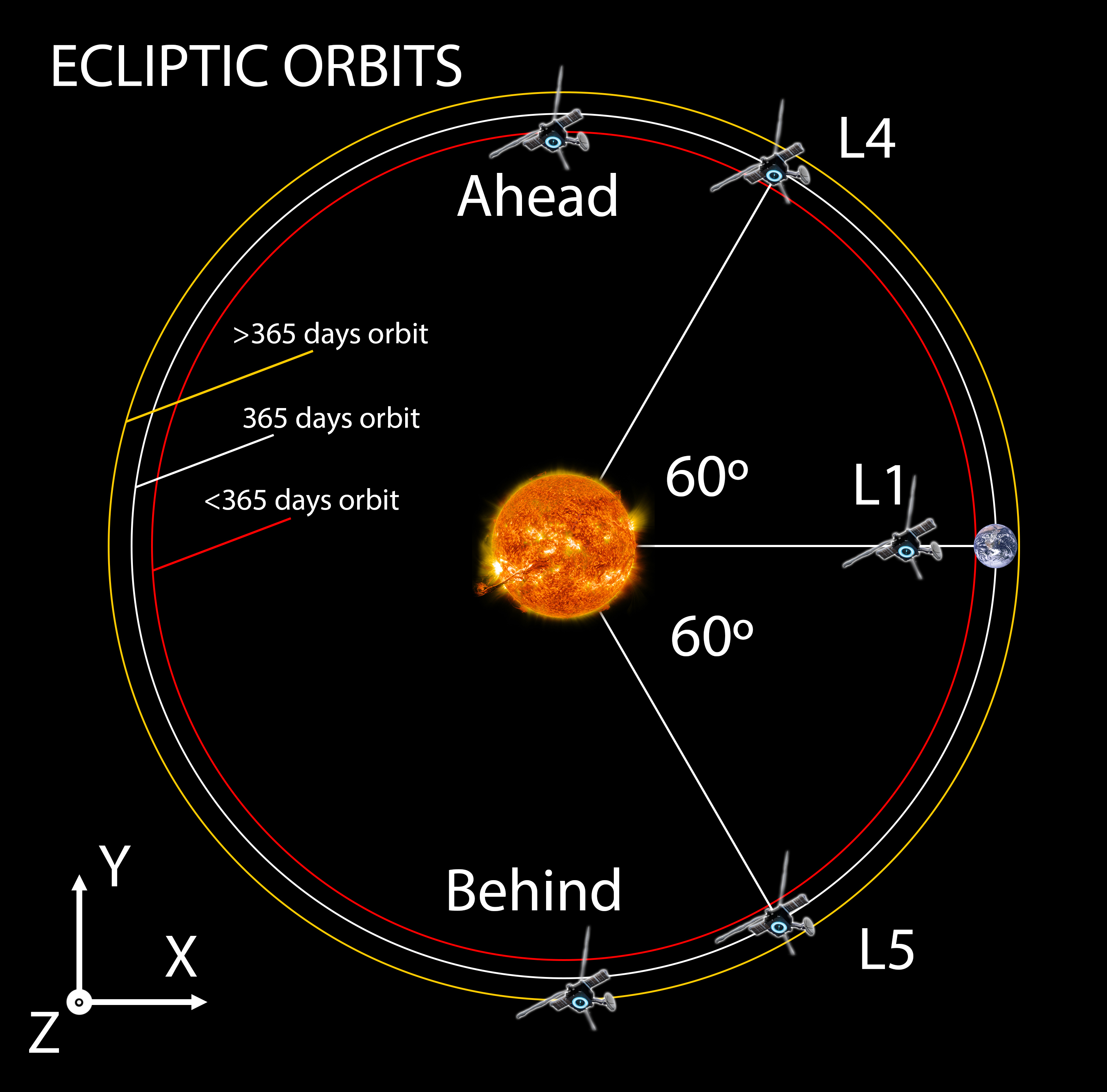}
	%	\noindent\makebox[\textwidth][c]{\includegraphics[width=1\textwidth]{photon_density_map_20001_paper_plot_v1.png}}%
	\caption{Final descoped configuration of the SURROUND mission concept composed of 5 CubeSats along the Sun-Earth ecliptic: three Lagrange spacecraft (L1, L4 and L5) and  two spacecraft deployed at orbits ahead and behind the Earth with a simultaneous drift rate.}
	\label{fig:orbits}
\end{figure}

In an effort to improve monitoring and forecasting capabilities in space weather observations, we propose the SURROUND concept mission, as part of the European Space Agency (ESA) Open Space Innovation Platform (OSIP) Nanosats for Spaceweather Campaign. In this proposal, a constellation of satellites in ecliptic orbit around the Sun (see Figure~\ref{fig:orbits}; 3 Lagrange spacecraft and 2 on an Earth-leading and lagging orbits similar to the Solar Terrestrial Relations Observatory [STEREO; \citet{eyles2009stereo}]) are used to provide co-temporal measurements that, when analysed using multilateration techniques (e.g., \citet{Bancroft1985tdoa,Smith1987tdoa}), will enable localisation and tracking of solar events. Through consideration of appropriate multilateration techniques and use cases, this paper defines suitable measurement requirements and summarises the key findings of the Phase-0 study. The Phase-0 is to effectively assess the feasibility of the SURROUND mission concept. Having multiple nanosatellites dedicated to operational monitoring of the heliosphere, for the first time,  allows us to improve current forecasting capabilities of these hazardous space weather phenomenon. This paper outlines the key scientific and mission requirements of the proposed SURROUND mission concept.

Current space-based solar observatories [e.g., Solar Orbiter (SolO; \citet{muller2013SolO}), Parker Solar Probe (PSP; \citet{fox2016psp})] and operational forecasting centres [e.g., STEREO, Geostationary Operational Environmental Satellites (GOES; \citet{goodman2013goes})] monitor and observe the solar surface and lower atmosphere over multiple wavelengths (i.e., white-light, extreme ultraviolet (EUV), X-rays); very few solely image the heliosphere. Therefore, although previous and ongoing campaigns have provided well documented catalogues of various space weather hazards (CME and flare onsets, solar wind evolution...), tracking the progress of eruptive events and thus forecasting the probabilistic nature of whether these events will impact the Earth's space environment still remains one of the greatest challenges in the field. Furthermore many of these phenomena are unobservable in EUV and white light wavelengths directly. For example, many CMEs can appear very faint in the lower corona and travel slower than typical events (i.e., `stealth CMEs'; e.g., \citet{OKane2019stealth} and references therein). These may be overlooked, hindering possible forecasts.  In addition, flare-associated  Solar Energetic Particles (SEPs) are by definition not accompanied by a discernable CME in coronagraph observations and do not emit and/or scatter radiation at EUV and white light wavelengths. However in both cases, associated electrons are observable from \textit{in situ} and solar radio measurements. This is particularly useful for SEP events which may be hazardous to spacecraft and communications. The lighter, less harmful electrons travel faster than the potentially hazardous heavier protons, providing a useful diagnostic to forecast such space weather events. These accelerated electrons appear as distinct features in dynamic spectra produced from radio spectrometry: Solar Radio Bursts (SRBs). 

\begin{figure}
	\centering
	\noindent\includegraphics[width=\textwidth]{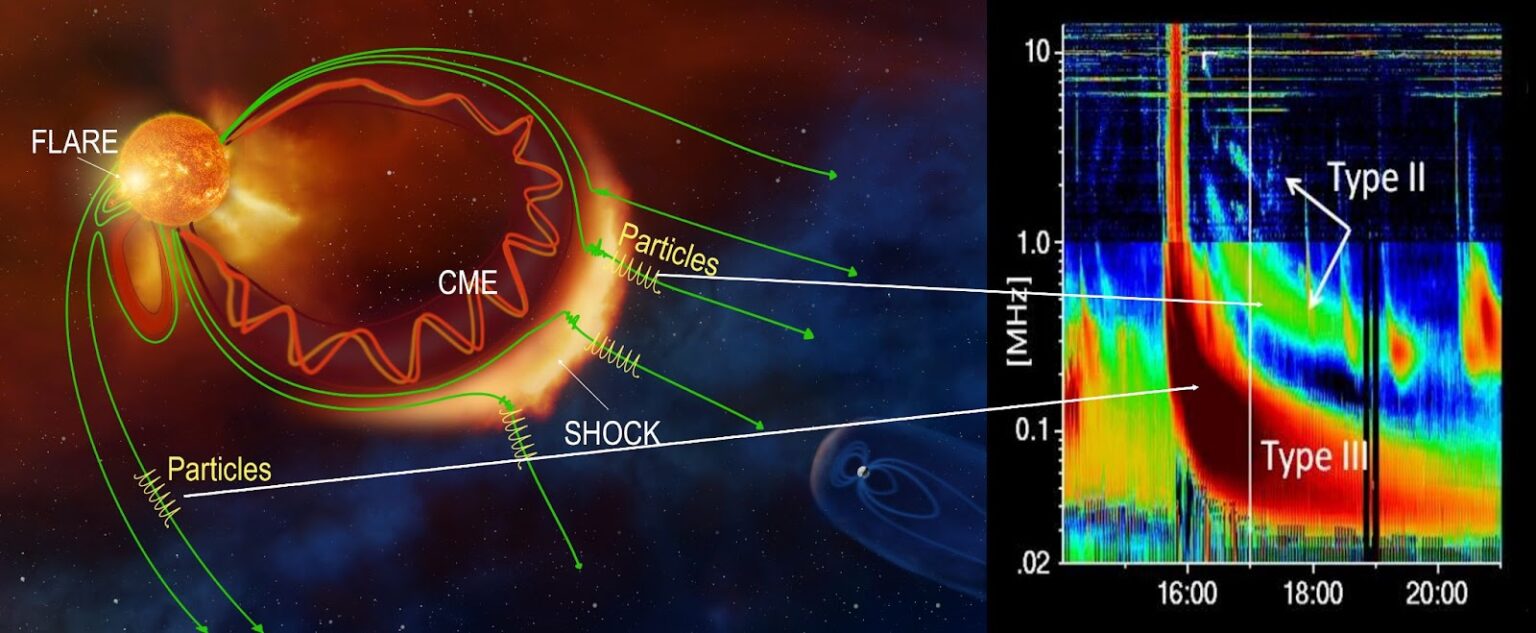}
	%	\noindent\makebox[\textwidth][c]{\includegraphics[width=1\textwidth]{photon_density_map_20001_paper_plot_v1.png}}%
	\caption{A (left) cartoon showing various eruptive events and (right) example spacecraft data (Wind/Waves data from \citet{gopalswamy2019srb}) of a radio dynamic spectra highlighting different spectral classes of SRBs. Type-II bursts are associated with the shock-front of a CME while Type-IIIs are linked with the acceleration of SEPs. Solar eruption illustration is adapted from ESA/A. Baker, CC BY-SA 3.0 IGO. \href{https://www.dias.ie/cosmicphysics/astrophysics/surround/}{Image source}.}
	\label{fig:srbs}
\end{figure}

SURROUND will focus on tracking \textit{Type-II} (associated with particle acceleration generated in the shock fronts of CMEs as they propagate through the interplanetary medium \citep{cane1982typeii,webb2012cmes}) and \textit{Type-III} (SEPs linked to electron beam acceleration travelling on open heliospheric magnetic field lines, generated from solar flare events \citep{reid2014typeiii}) spectral types, over a frequency range of \qtyrange{0.02}{25}{\mega\hertz} (i.e., low frequency solar radio emissions). Both can be identified from dynamic spectra produced from radio spectrometry over this frequency range (see Figure~\ref{fig:srbs}). Space weather monitoring facilities use Type-II and Type-III SRBs as diagnostics for incoming CMEs and accelerated SEPs respectively. With this, the goal of space weather operations is to provide accurate forecasting/nowcasting information within the propagation times of these space weather events, using the radio emissions from accelerated electrons as our diagnostics: within 12 hours to $\sim$ 9 days for CMEs \citep{webb2012cmes}, and $\sim$ \qtyrange{21}{28}{\min} for the lighter electrons from SEP events with the heavier protons following $\sim$ a few hours behind (e.g., \citet{reames2013seps, reid2014typeiii}). With multiple viewpoints to track any SRBs travelling towards Earth, SURROUND will be the flagship for tracking and monitoring space weather.

%The remainder of this work will discuss the science behind the methods SURROUND will employ to track SRBs (Section~\ref{sec: srbs}) and the finalised Mission Requirements of this Phase-0 study (Section~\ref{sec: mrs}). The final section, Section~\ref{sec: outcome}, will briefly outline the final outcomes of the Phase-0 study driven by the Mission Requirements and both the Mission and System analysis and the next stages in the process.

\section{Tracking Solar Radio Bursts (SRBs)}\label{sec: srbs}

The SURROUND mission is designed to track Type-II and -III SRBs using multilateration techniques. Such methods are designed for multiple receivers to provide information on the location and approximate size of the source of interest.

\subsection{Modelling the best case scenario}
The easiest multilateration technique to implement is the Time-of-Arrival (TOA) method which is reliant on temporal knowledge the emission of the source signal. The distance from the source can be determined from the time difference between the time of arrival ($t_{\mathrm{arrival}}$) and the time of emission ($t_{\mathrm{emission}}$), assuming a constant propagation velocity $c$ (for radio emissions, this is the speed of light in a vacuum: $c$ $\approx\ 3.00 \times 10^{8}$\,ms$^{-1}$):

\begin{equation}\label{eq:toa}
    d = c(t_{\mathrm{arrival}} - t_{\mathrm{emission}})
\end{equation}

\begin{figure}
	\centering
	\noindent\includegraphics[width=\textwidth]{ 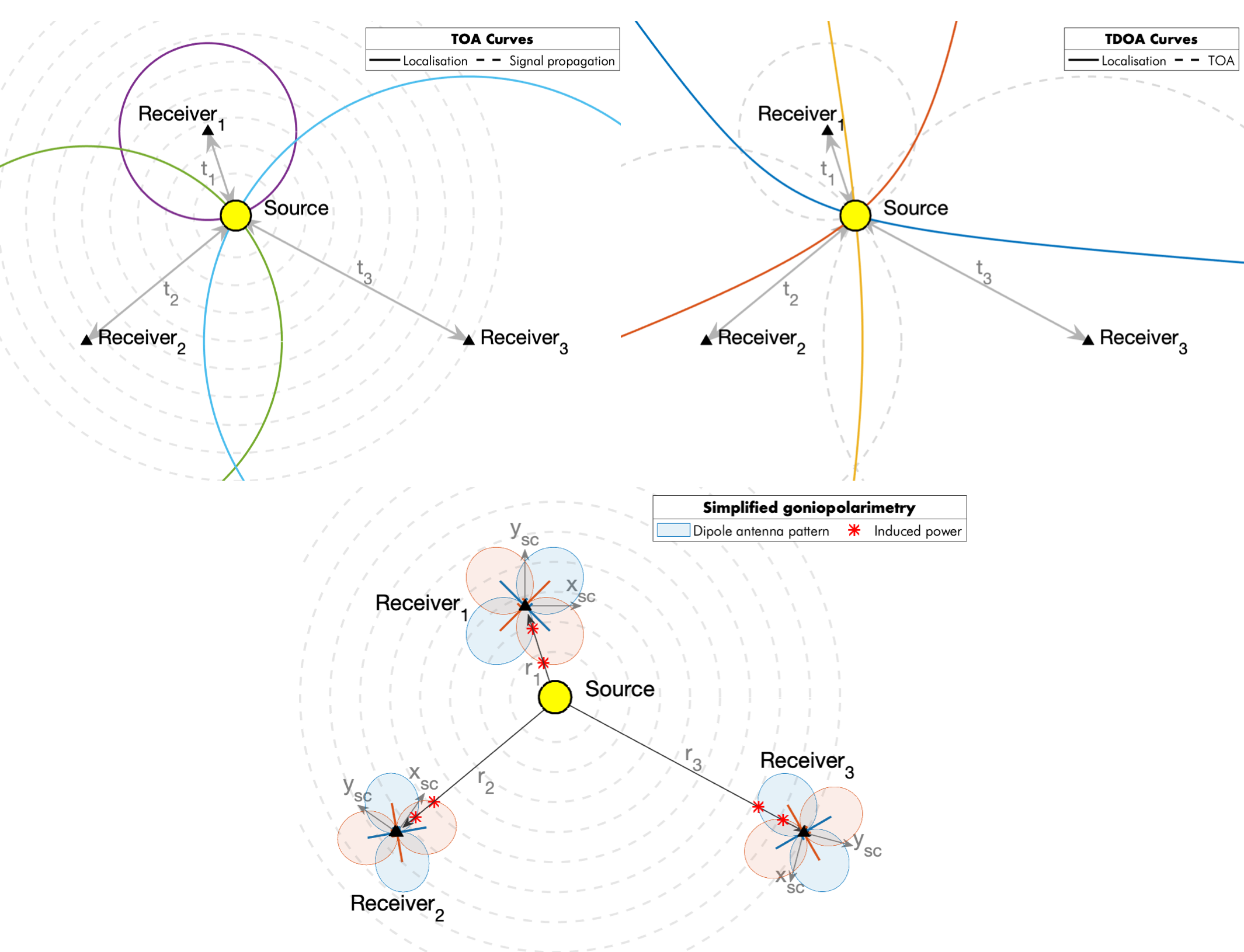}
	\caption{Schematic of how both the (top-left) TOA  and (top-right) TDOA techniques determine the location of a given source. Source location is determined from the intersection of circumferences and hyperbola pairs for TOA and TDOA respectively. Bottom panel shows a simplified cartoon of the goniopolarimetry (GP) method with antenna beam pattern and resultant reconstructed wave power back-propagated to source (red).}
	\label{fig:multi}
\end{figure}

For a single receiver this corresponds to a circle with radius length $d$ of which the circumference defines all possible source locations detected by the particular receiver, as shown in Figure~\ref{fig:multi}. The location of the source can be solved by use of parametric equations. If there are at least three receivers available, 2D multilateration can be performed (similarly four receivers is needed for the three-dimensional case). This is given by the following equation of the circle
\begin{equation}
    d_{r}^{2} = (x_{r}-x)^{2} + (y_{r} - y)^{2}
\end{equation}

where $d_{r}$ and ($x_{r}, y_{r}$) are the distance from the source and positional 2D coordinates of each individual receiver and ($x,y$) are the source coordinates.

For the case of SURROUND, the spatial and temporal information of an emitted SRB is unknown and therefore this technique is not practical from an operations perspective. However, TOA can be used to describe the ``best case'' scenario for tracking these bursts with very close overall performance to more complex methods (e.g., \citet{kaune2012accuracy}). Here we use a fully geometrical TOA method as an ideal case to compare with a Bayesian algorithm where the probabilistic position of the SRB or posterior distribution, $P(x,v|\Delta t)$, is calculated given that there is an observing difference in time $\Delta t$ between source and reciever without any given knowledge of the signal wave propagation speed (e.g., \citet{salvatier2016pymc3}). This is described by a version of Bayes' Theorem considering three parameters:

\begin{equation}\label{eq: bayes}
    P(x,v|\Delta t) = \frac{P(\Delta t|x,v)P(x|v)P(v)}{P(\Delta t)}
\end{equation}

where $x$ and $v$ are the source position and wave propagation velocity. The data likelihood, $P(\Delta t|x,v)$, is the physics model used in the algorithm - for the case of SURROUND we modelled the source-receiver distance from each satellite with added noise to simulate possible propagation effects. The prior distributions for the given parameters ($x, v$ and $\Delta t$) are given by $P(x|v), P(v)$ and $P(\Delta t)$ where $P(x|v)$ denotes the prior distribution of the source position given the propagation velocity of the source (or SRB). 

\subsection{Multilateration and direction finding with SURROUND}

The time-difference-of-arrival (TDOA) multilateration technique is more complex than TOA as it relies on the difference between the times of arrival of the signal at each receiver (or satellite) to localise the source. Therefore the TDOA method is independent of the time of emission and more applicable to the scenarios SURROUND will encounter. As a result, we obtain a difference in distance between the receivers to the source:

\begin{equation}
    \Delta d_{i,j} = c(t_{i} - t_{j}) = \sqrt{\left(x_{j} - x\right)^{2} + \left(y_{j} - y\right)^{2}} - \sqrt{\left(x_{i} - x\right)^{2} + \left(y_{i} - y\right)^{2}}
\end{equation}

where $t_{i,j}$ is the time of arrival to receivers $i$ and $j$; ($x_{i,j}, y_{i,j}$) are the known positions of receivers $i$ and $j$ and ($x,y$) is the hyperbolic solution for the position of the source. From these solutions, TDOA uses multiple satellites to isolate an area where the likelihood of the source location is high (as shown in Figure~\ref{fig:multi}). Therefore increasing the number of satellites and deploying them in a calculated, strategic position around the source emission will decrease the error in source location (i.e., the area of intersection between pairs of hyperbola decreases), given the separation angle is large enough. For multiple radio sources (e.g. along a CME shock front) the calculated TDOA will be different and therefore in a different position in the inner heliosphere. To disentangle these sources, we can trace along a distinctive feature of the SRB on the dynamic spectra by fitting a line profile (e.g., leading edge for Type-III bursts) to find the TOA at each satellite and use multilateration to determine the position across multiple frequency channels. 

The main assumption used in both of TDOA and TOA is that the measurements are based on the shortest target-receiver path the signal can take, or the Line of Sight. The signal will be very weak or unobserved if this path is blocked, leading to erroneous results in the location or possibly very large uncertainties.

%within the inner heliosphere. However, such scattering effects can alter the spectral features of radio bursts and can alter their appearance on dynamic spectra \citep{kontar2019anisotropic} and therefore should be accounted for. 

SURROUND will also be capable to fully implement the goniopolarimetry (GP) method (numerical solutions found in \citet{Cecconi2005}). GP functions by determining the propagation vector of detected electromagnetic waves emitted from the source using three calibrated and carefully positioned antennae (or two with one acting as a spinning dipole). Each antenna (or dipole antenna) measures the polarisation and flux of the detected radio emissions which are then used to reconstruct the resulting wave, and back propagate to the source within a systematic uncertainty (dependent on parameters of the radio antenna; see simple configuration in Figure~\ref{fig:multi}). For the case of SURROUND, using GP with multiple CubeSats observing radio signals from the same source, will result in each individual wave vector being back propagated to the same position. Where these wave vectors intersect will provide the location of the radio source when observed at the same frequency across all spacecraft (required to perform GP). Full modelling and more rigorous analysis of the GP method will be carried out in the next mission phase.

This method has been implemented successfully from using a single spacecraft (e.g., analysis of Saturn's kilometric radiation from Cassini \citep{cecconi2009skr}) and  multiple spacecraft (e.g., tracking and triangulating a Type-III burst using Wind and Ulysses \citep{reiner1998winduly}); both methods treating the source as point-like. For the case of SRBs however, due to scattering in the interplanetary medium, the SRBs will appear spatially extended \citep{Cecconi2007extend} and therefore the original GP treatment from \citet{Cecconi2005} need to account for this. This has been implemented in previous studies tracking Type-II (e.g., \citet{Oliveros2012typeii}) and -III SRBs (e.g., \citet{Krupar2012}). Therefore the expertise on calibrating the antenna and analytically solving the equations dealing with the correlations between all three antennae for an extended source is all readily available.

In the case of SURROUND, GP capable antennas would be beneficial to the mission as it will allow SRB localisation and tracking to be performed at small separation angles when TOA/TDOA is very inaccurate and/or fails. Additional information of the radio bursts (e.g., polarisation) detected from a 3-antenna setup will be useful for further science cases (see Section~\ref{sec: mrs}). Given the low directivity of the Type-II and -III emissions we are interested in, the errors produced from multilateration are minimised and the Poynting vectors for GP (e.g., resulting wave vectors) are not parallel when the satellites are at varied distances from the Earth with a reasonable separation angle. Therefore these methods take advantage of the low directivity to improve their performance.

{As solving the GP equations is more complex than the TOA/TDOA methods, this is computationally more expensive. Therefore for providing forecasts within ~30-min (see Section~\ref{sec: mrs}) the multilateration methods will be preferable. However, the same antenna configuration can be applied to both sets of methods reducing the complexity of our payload. We do note that these methods for tracking SRBs initially assume that the radio signal produced from the source is isotropic and is unaffected by inharmonious scattering along the signal path, and that any errors will be dominated by the uncertainty in source location. Although, as noted in previous studies (e.g., \citet{Bonnin2008}), the effective beam of the source (electron beam with scattering effects) is significantly anisotropic and will affect our source location. Such effects will be explored and analysed in detail in the later stages of the mission after Phase-0 (i.e., Phase-A).

\section{Mission Requirements}\label{sec: mrs}

The science goals of SURROUND can be summarised in the following statement: \textit{``The SURROUND mission will detect, localise and track solar radio bursts in 3D using a constellation of radio spectrometers and to forecast the arrival of the radio bursts' drivers at Earth (or Earth-like distances)."}

The aim of this section is to synthesise the above mission statement and briefly explain the drivers and requirements motivating the SURROUND concept. Here we start by explaining the Use Cases of the mission; preliminary results and outcomes for the use of multilateration techniques to track SRBs and then conclude by discussing the Measurement requirements needed to fulfil the science goals and drive the design of the mission and spacecraft.

\subsection{Use Cases}
The SURROUND mission will focus on space weather operations and therefore the primary use case of the concept mission is the detection, tracking and localisation of SRBs. This will provide us with a novel and more accurate way of forecasting the arrival of these events at the Earth. As discussed throughout this Phase-0 study, SURROUND will primarily focus on two types of SRBs: (i) Type-II and (ii) Type-III bursts. SURROUND's main Use Cases (UCs) focus on detecting and tracking SRBs (Table~\ref{tab:uses}: UCs 1 and 2); localise and determine the location of SRBs (UCs 3 and 4); providing nowcasting results of CMEs and SEP electrons (UCs 5 and 6) and forecast the arrival of space weather drivers within their propagation timescales (UCs 7-9). These Use Cases are summarised in Table~\ref{tab:uses} and will drive the Measurement Requirements to determine the specifications of the payload needed (e.g., length of antenna, frequency range, sensitivity...).

\begin{table*}[htp]
    \scriptsize
    \begin{center}
        \caption{Summary of SURROUND's Use Cases}
        \label{tab:uses}
        \begin{tabular}{|l|l|l}
            \cline{1-2}
            \multicolumn{1}{|c|}{\textbf{\begin{tabular}[c]{@{}c@{}}Use Case\\ (UC)\end{tabular}}} & \textbf{Description}                                                                                                                                     &  \\ \cline{1-2}
            \textbf{UC1}                                                                         & Detect and track Type-II radio bursts - fast CMEs shocks accelerated electron beams                                                                      &  \\ \cline{1-2}
            \textbf{UC2}                                                                         & Detect and track Type-III radio bursts - flare accelerated electron beams                                                                                &  \\ \cline{1-2}
            \textbf{UC3}                                                                         & Localise in 3D Type-II radio bursts                                                                                                                      &  \\ \cline{1-2}
            \textbf{UC4}                                                                         & Localise in 3D Type-III radio bursts                                                                                                                     &  \\ \cline{1-2}
            \textbf{UC5}                                                                         & Nowcast CMEs associated with Type-IIs                                                                                                                    &  \\ \cline{1-2}
            \textbf{UC6}                                                                         & Nowcast SEP electrons associated with Type-IIIs                                                                                                          &  \\ \cline{1-2}
            \textbf{UC7}                                                                         & \begin{tabular}[c]{@{}l@{}}Forecast arrival of CMEs associated with Type-II at Earth* (propagation time \\ scale $\sim$12 hours)\end{tabular}             &  \\ \cline{1-2}
            \textbf{UC8}                                                                         & \begin{tabular}[c]{@{}l@{}}Forecast arrival of SEP electrons associated with Type-III at Earth* (propagation \\ time scale $\sim$10 minutes)\end{tabular} &  \\ \cline{1-2}
            \textbf{UC9}                                                                         & \begin{tabular}[c]{@{}l@{}}Forecast the potential arrival of SEP protons associated with Type-III \\ (propagation time scale $\sim$2 hours)\end{tabular} &  \\ \cline{1-2}
            \textbf{UC10$^\dagger$}                                                                        & Scientific Analysis                                                                                                                                      &  \\ \cline{1-2}
            \multicolumn{3}{l}{$^{*}$ or at Earth like distances} \\
            \multicolumn{3}{l}{$^{\dagger}$ SURROUND's secondary Use Case} \\

        \end{tabular}

    \end{center}
\end{table*}

The secondary Use Case proposed for SURROUND is scientific analysis of obtained mission data. While the primary objectives focus on providing operational information for space weather monitoring and forecasting, the gathered data will be beneficial for many in the solar and planetary science community. For example, a minimum of three different viewpoints of the solar radio emissions, all providing simultaneous measurements, allows the Sun's global behaviour to be observed in finer detail when compared to current observatories. The multilateration techniques can also incorporate other solar/heliospheric monitors further improving the accuracy of SURROUND's localisation and tracking measurements (e.g., further reducing the area of intersection). Utilising these combined results, or solely SURROUND measurements, will improve the potential now/fore-casting capabilities of the mission itself.

\subsection{Measurement Requirements}

Using the Use Cases as a foundation (Table~\ref{tab:uses}), we identify the main requirements essential for the SURROUND mission to achieve its science objectives. These are outlined in Table~\ref{tab:meas} with the corresponding numbered Use Case(s) that influenced the noted requirement and are essential to the next stages of the mission analysis. Although the main focus of this paper is more the science aspect of the SURROUND mission, these stages, together with the Mission Requirements, will determine the final outcomes of this Phase-0 Study (as outlined in Section~\ref{sec: outcome}). 

\begin{table}[htp]
    \scriptsize
    \begin{center}
        \caption{Summary of SURROUND's Measurement Requirements}
        \label{tab:meas}
        \begin{tabular}{|l|l|l|l|l}
            \cline{1-4}
            \textbf{\begin{tabular}[c]{@{}l@{}}MR*\end{tabular}} & \textbf{Description}                                                                                                                                                                                                                                                          & \textbf{Key Driver}                                                                                                                                                                      & \textbf{\begin{tabular}[c]{@{}l@{}}Use\\ Cases\end{tabular}} &  \\ \cline{1-4}
            \textbf{MR1}                                                                    & \begin{tabular}[c]{@{}l@{}}Detect radio waves with spectral\\  flux density of greater than \\ \qty{10e-22}{\W\per\m\tothe{-2}\hertz\tothe{-1}}\end{tabular} & \begin{tabular}[c]{@{}l@{}}Detection of Type-II and \\ Type-III radio bursts\end{tabular}                                                                                                & \begin{tabular}[c]{@{}l@{}}UC1,\\ UC2\end{tabular}           &  \\ \cline{1-4}
            \textbf{MR2}                                                                    & \begin{tabular}[c]{@{}l@{}}Detect radio waves with in frequency \\ range of \qtyrange{0.02}{25}{\mega\hertz}\end{tabular}                                                                                                  & \begin{tabular}[c]{@{}l@{}}Track radio burst from close to \\ solar ($\sim$2\,$R_{\odot}$ surface \\ to \qty{1}{\astronomicalunit}\end{tabular} & \begin{tabular}[c]{@{}l@{}}UC1,\\ UC2\end{tabular}           &  \\ \cline{1-4}
            \textbf{MR3}                                                                    & \begin{tabular}[c]{@{}l@{}}Scan the frequency range of \\ \qtyrange{0.02}{25}{\mega\hertz} \\ at least every \qty{10}{\s}\end{tabular}                                                      & \begin{tabular}[c]{@{}l@{}}Achieve accurate 3D localisation \\ using TOA/TDOA methods\end{tabular}                                                                                       & \begin{tabular}[c]{@{}l@{}}UC1,\\ UC2\end{tabular}           &  \\ \cline{1-4}
            \textbf{MR4}                                                                    & \begin{tabular}[c]{@{}l@{}}Minimum of 3 observation \\ locations (e.g., $L_1$,$L_4$, $L_5$) with \\ separation of at least \ang{50}\end{tabular}                                                                                                             & \begin{tabular}[c]{@{}l@{}}TOA/TDOA methods require at \\ least 3 different viewpoints (GP \\ has minimum of 2)\end{tabular}                                                             & \begin{tabular}[c]{@{}l@{}}UC3,\\ UC4\end{tabular}           &  \\ \cline{1-4}
            \textbf{MR5}                                                                    & \begin{tabular}[c]{@{}l@{}}Obtain data from spacecraft at least \\ every 6 hours\end{tabular}                                                                                                                                                                                 & \begin{tabular}[c]{@{}l@{}}Required time to process, generate \\ and issue nowcasts for Type-IIs\end{tabular}                                                                            & UC5                                                          &  \\ \cline{1-4}
            \textbf{MR6}                                                                    & \begin{tabular}[c]{@{}l@{}}Obtain data from spacecraft at least \\ every 15 minutes\end{tabular}                                                                                                                                                                              & \begin{tabular}[c]{@{}l@{}}Required time to process, generate \\ and issue nowcasts for Type-IIIs\end{tabular}                                                                           & UC6                                                          &  \\ \cline{1-4}
            \textbf{MR7}                                                                    & \begin{tabular}[c]{@{}l@{}}Obtain data from spacecraft at least \\ every 3 hours\end{tabular}                                                                                                                                                                                 & \begin{tabular}[c]{@{}l@{}}Required time to process, generate \\ and issue forecasts for CMEs \\ associated with Type-IIs\end{tabular}                                                   & UC7                                                          &  \\ \cline{1-4}
            \textbf{MR8}                                                                    & \begin{tabular}[c]{@{}l@{}}Obtain data from spacecraft at least \\ every 10 minutes\end{tabular}                                                                                                                                                                              & \begin{tabular}[c]{@{}l@{}}Required time to process, generate \\ and issue forecasts for SEP electrons \\ associated with Type-IIIs\end{tabular}                                         & UC8                                                          &  \\ \cline{1-4}
            \textbf{MR9}                                                                    & \begin{tabular}[c]{@{}l@{}}Obtain data from spacecraft at least \\ every hour\end{tabular}                                                                                                                                                                                    & \begin{tabular}[c]{@{}l@{}}Required time to process, generate \\ and issue forecasts for potential SEP \\ protons associated with Type-IIIs\end{tabular}                                 & UC9                                                          &  \\ \cline{1-4}

            \multicolumn{5}{l}{*MR = Measurement Requirement}
        \end{tabular}

    \end{center}
\end{table}

The mission requirements can be grouped into four categories which we define and briefly expand upon here:

\begin{enumerate}
    \item \textit{\textbf{Frequency range and sensitivity (MR1, MR2):}} The frequency range of the antennas (needed for all multilateration methods) is calculated from the distances SURROUND will be able to detect, track, localise and forecast SRBs using plasma density models \citep{Leblanc1998,parker1958dynamics,saito1977study}. These density models suggest that SURROUND will be capable of tracking bursts from 2\,$R_{\odot}$ (or 2\,$R_\mathrm{Sun}$: near the solar surface) to 215\,$R_{\odot}$ (Earth's orbit) corresponding to a frequency range of \qtyrange{0.2}{25}{\mega\hertz}. The sensitivity of SURROUND's antennas should be such that radio waves with a spectra flux density exceeding \qty{10e-22}{\W\per\m\tothe{-2}\hertz\tothe{-1}} should be detectable. Both frequency and sensitivity specifications are feasible for SURROUND and have been successfully implemented on previous missions (i.e., plasma and radio instrument on SolO \citep{maksimovic_solar_2020}). The capability of tracking SRBs to 215\,$R_{\odot}$ will enable us to provide multiple data of time evolution of the electrons as they propagate through the inner heliosphere for scientific analysis (secondary Use Case) to obtain useful diagnostics of the source plasma (i.e. temperature, density profile).

 %    \begin{figure}
	% \centering
	% \noindent\includegraphics[width=0.6\textwidth]{Figures/DensityModels.png}
	% %	\noindent\makebox[\textwidth][c]{\includegraphics[width=1\textwidth]{photon_density_map_20001_paper_plot_v1.png}}%
	% \caption{Schematic of how both the (left) TOA  and (right) TDOA techniques determine the location of a given source. Source location is determined from the intersection of circumferences and hyperbola pairs for TOA and TDOA respectively.}
	% \label{fig:desnity}
 % \end{figure}

    \item \textit{\textbf{Time resolution (MR2, MR3):}} The time cadence of the spectrometers used in SURROUND is essential in order to obtain the best quality data at an optimised resolution. Using multilateration techniques also introduces a time resolution-dependent error on the location of the radio source with time cadence (the higher the cadence, the smaller the error on location). However the higher time cadence leads to a larger volume of data that needs to be processed and transmitted back to ground and therefore also needs to be considered when addressing the data resolution. For a SURROUND mission consisting of 5 spacecraft in the ecliptic plane, a \qty{10}{\second} cadence is desirable for all spectrometers. These data files can be transmitted to the ground receiver during days of high solar activity and increased likelihood of solar flaring. Quieter times of less priority will be measured using a 60s cadence (i.e., lower time resolution). Similar results were also obtained for a 4-spacecraft mission in the event of one CubeSat being lost. 
    
    \item \textit{\textbf{Spacecraft orbits and position (MR4):}} In order to achieve the science objectives and maintain sufficient accuracy for tracking and localising SRBs, SURROUND will require at least 3 - 5 satellites. As demonstrated by the initial spacecraft configuration shown in Figure~\ref{fig:orbits}, each satellite will be placed in strategic orbits to maintain sufficient angular separation to perform TOA/TDOA and optimised viewing of the Earth-facing of the Sun at all times. 
    
    \item \textit{\textbf{Frequency of downlinks (MR5 - MR9):}} The frequency of downlinks depends on the type of space weather phenomena being monitored depending on their Sun-Earth travel time. For example, radio emissions produced from Type-III SRBs will take $\sim$ \qty{8}{\min} to reach the SURROUND spacecraft. The associated electrons, travelling at $\sim$ \qty{0.3}{c}, will reach Earth in $\sim$ \qty{30}{\min}. Current capabilities can allow us to assume data processing and transmission will take  \qty{5}{\min} (e.g., $\sim$ 15-min buffer to generate forecast for electrons). Therefore data acquisition needs to be at a higher frequency (e.g., 10-min). The frequency has to be optimised to allow the burst to travel far enough to derive propagation direction but fast enough to enable the data to be processed and issue a forecast well below the typical propagation time (see Table~\ref{tab:meas}). The size of the data products to be communicated to the ground receiver will also affect the downlink time. For SURROUND, we estimate each antenna will have a daily data rate of $\sim$ \qtyrange{120}{150}{MB} (depending on time cadence) in broad agreement with the radio and plasma waves (RPW) instrument onboard SolO (assuming a 256 spectral channel spectrometer with 12-bit resolution and 35\% margin accounting for overheads and general housekeeping of the data).
\end{enumerate}

\section{Outcomes of SURROUND Phase-0 study}\label{sec: outcome}
Based on the results from the work presented here and various discussions and feasibility studies we can draw our final outcomes of this Phase-0 study including, but not limited to:

\begin{itemize}
    \item SURROUND can achieve its mission objectives using either a 3, 4 or 5-spacecraft  (in order of highest uncertainty of SRB localisation).
    \item Deployment to $L_4$ and $L_5$ are the most challenging and requires most propulsion. $L_1$ spacecraft, closest to Earth, requires orbit maintenance over mission lifetime.   
    \item Tracking and localisation of various SRBs can be carried out using multilateration and GP techniques, using the same antenna configuration.
    \item With current technological and ground communication constraints (need to retrieve from \qty{1}{\astronomicalunit}), data can be preferentially received at high cadence from $L_1$ (every 30-mins to 1-hr) and at longer regular intervals from the remainder of the constellation (at lower cadence).
    \item Forecasting potential SEP and incoming CMEs is feasible (time cadence dependent) given current technologies. 
\end{itemize}

The results of this Phase-0 study suggest that the SURROUND is feasible and providing potential solutions for the more challenging obstacles (e.g., spacecraft deployment/maintenance, data downlink timings for forecasts). There is no other space weather monitoring mission (currently ongoing or in advanced planning stages) like SURROUND, which will revolutionise how we track space weather and how we can utilise CubeSats for deep space science.

\section*{\small Acknowledgements}
\footnotesize The SURROUND phase-0 study is funded by European Union's Horizon 2020 research and innovation programme under Grant agreement No. 952439 and project number AO 2-1927/22/NL/GLC/ov as part of the ESA OSIP Nanosats for Spaceweather Campaign
\newcommand{\newblock}{}
\bibliographystyle{mnras}
\scriptsize
%%%%%%%%%%%%%%%%
\bibliography{surround.bib}

\end{document}